\pdfoutput=1 
\documentclass{nature}
\usepackage{underscore}
\usepackage{multirow}
\usepackage{ccaption}
\usepackage{subcaption}
\usepackage{graphicx}
\usepackage{lipsum}
\usepackage{ragged2e}
\usepackage{fixltx2e}
\usepackage{hyperref}

\usepackage{cmbright}
\usepackage{filecontents}
\usepackage{amsmath}
\renewcommand\refname{References}

\newcounter{lastnote}

\usepackage{comment}
\usepackage{cite}
\usepackage{adjustbox} 

\makeatletter
\newenvironment{figurehere}
{\def\@captype{figure}}
{}
\makeatletter

\usepackage{xcolor}
\definecolor{sared}{rgb}{0.69, 0,0}
\title{\begin{center}  A Full Spectrum of 3D Ferroelectric Memory \\ Architectures Shaped by Polarization Sensing\end{center}}

\author
{ \centering
{
\large{Jiahui Duan$^{1\dag}$, Asif Khan$^{2}$, Xiao Gong$^{3}$, Vijaykrishnan Narayanan$^{4}$, Kai Ni$^{1}$\\}
\vspace{3ex}
\normalsize{$^{1}$University of Notre Dame, Notre Dame, IN 46556, USA}\\
\normalsize{$^{2}$Georgia Institute of Technology, Atlanta, GA 30332, USA;}\\
\normalsize{$^{3}$National University of Singapore, Singapore;}\\
\normalsize{$^{4}$Pennsylvania State University, State College, PA 16802, USA}\\
\vspace{2ex}
\normalsize{$^{\dag}$To whom correspondence should be addressed} \\
\normalsize{Email: jduan3@nd.edu} \\
}}

\begin{document}
\flushbottom
\maketitle
\vspace{3ex}
\begin{abstract}
Ferroelectric memories have attracted significant interest due to their non-volatile storage, energy efficiency, and fast operation, making them prime candidates for future memory technologies. As commercial Dynamic Random Access Memory (DRAM) and NAND flash memory are transiting or have moved toward three-dimensional (3D) integration, 3D ferroelectric memory architectures are also emerging, provided they can achieve a competitive position within the modern memory hierarchy. Given the excellent scalability of ferroelectric HfO\textsubscript{2}, various dense 3D integrated ferroelectric memory architectures are feasible, each offering unique strengths and facing distinct challenges.
In this work, we present a comprehensive classification of 3D ferroelectric memory architectures based on polarization sensing methods, highlighting their critical role in shaping memory cell design and operational efficiency. Through a systematic evaluation of these architectures, we develop a unified framework to assess their advantages and trade-offs. This classification not only enhances the understanding of current 3D ferroelectric memory technologies but also lays the foundation for designing next-generation architectures optimized for advanced computing and high-performance applications.
\end{abstract}

%
%
\thispagestyle{empty}


\section*{\textcolor{sared}{\large Introduction}}

Ferroelectric memory is a type of non-volatile memory that uses ferroelectric polarization to store data. Unlike conventional memory technologies such as DRAM, which rely on temporary charges stored in capacitors using linear dielectric materials, ferroelectric memory stores data using the stable polarization states of ferroelectric material\cite{schenk2020memory,schenk2021new,mikolajick2020past}. This allows it to retain information even when power is turned off, making it non-volatile. It is also known for its fast read and write speeds, low power consumption, and high endurance, making it a prime candidate as an emerging memory to break the memory wall\cite{li2020reproducible}. Traditional ferroelectric memory, based on perovskite materials such as lead zirconate titanate (PZT), has long been investigated for its high-speed operation and low power consumption\cite{shirane1952phase,buck1952ferroelectrics}. However, its widespread adoption was hindered by integration challenges as a result of material incompatibility with modern complementary metal-oxide-semiconductor (CMOS) processes and also degradation of ferroelectric properties with scaled-down thickness\cite{junquera2003critical,liu2005thickness}. After the discovery of the ferroelectricity of the orthorhombic phase HfO\textsubscript{2} in 2011\cite{boscke2011ferroelectricity}, ferroelectric memory has gained significant attention as a promising candidate for next-generation non-volatile memory technologies\cite{muller2012ferroelectricity,mulaosmanovic2015evidence,florent2018vertical,bae2020sub,park2018review,kim2021ferroelectric,schenk2020memory,mulaosmanovic2021ferroelectric,jiao2023ferroelectric,chen2022hfo2}. The emergence of HfO\textsubscript{2}-based ferroelectric materials has reinvigorated interest in ferroelectric memory, offering a manufacturable solution compatible with advanced semiconductor fabrication\cite{hsain2022many}. Furthermore, HfO\textsubscript{2}-based ferroelectric materials can be integrated into ultra-thin film structures, allowing for aggressive scaling while maintaining ferroelectric properties\cite{xu2016general,kim2019ferroelectric,cai2023hzo,cai2024understanding,cheema2020enhanced,lancaster2024thickness,khan2020future}, making them highly suitable for advanced technology nodes.

The polarization sensing mechanism of the ferroelectric memory plays a critical role in determining the sense margins, cell structures, and scalability. To further enhance memory density, adopting a 3D array structure becomes essential. The sensing mechanism also deeply influences the 3D integration. In this review, we provide a comprehensive and systematic classification of 3D ferroelectric memory architectures based on their polarization sensing mechanisms, exploring their impact on memory cell design, performance, and scalability. By evaluating a wide range of architectures within this framework, we summarize their advantages, challenges, trade-offs in sense margin and scalability, and integration feasibility. Additionally, we discuss the prospects of ferroelectric memory applications. This classification not only deepens our understanding of emerging 3D ferroelectric memory technologies but also offers valuable insights into the design optimizations and sensing improvements required for future advancements. Furthermore, our framework serves as a foundation for developing next-generation memory architectures tailored for high-performance, low-power, and AI-driven computing applications, paving the way for breakthroughs in ferroelectric memory technology.

\section*{\textcolor{sared}{\large Ferroelectric Memory Classification}}
 
Fig. \ref{fig:Ferroelectric Memory Classification} provides an overview of various ferroelectric memory structures, categorized by their sensing mechanisms, along with a qualitative comparison of key performance metrics such as write voltage, retention, write and read endurance, sense margin, and scalability. The choice of polarization sensing mechanism influences the memory cell structure, array organization, and 3D integration strategy, resulting in a diverse range of ferroelectric memories, each with distinct advantages and challenges. Before exploring individual sensing operations in detail, we first introduce each type of memory.
The 1T-1C FeRAM consists of a single access transistor and a ferroelectric metal-ferroelectric-metal (MFM) capacitor stacked on the transistor's source or drain, closely resembling the conventional 1T-1C DRAM structure. However, unlike DRAM, which relies on volatile charge stored on the plate of a metal-insulator-metal (MIM) capacitor, FeRAM leverages the remanent polarization of the ferroelectric film to store data, offering non-volatility and lower power consumption. Since the write voltage is dropped solely across the MFM capacitor, it can be kept low and reduced to below 1V with HfO\textsubscript{2} film\cite{wang2023stable,sung2021low,lee2024beol}. Additionally, the MFM capacitor in the FeRAM inherits good endurance\cite{muller2013ferroelectric,lee2022investigation,jang2024demonstration,okuno2020soc,francois2019demonstration,popovici2022high,gong202110,walke2024doped} (\textgreater10\textsuperscript{12} cycles) and excellent retention\cite{wu20239} (\textgreater10 yrs) benefiting from clean interfaces and a low depolarization field due to metal electrode screening. The MFM polarization can be sensed in two main ways: direct charge sensing via destructive switching\cite{francois202116kbit,laguerre2023memory} or nondestructive capacitive sensing\cite{mukherjee2023capacitive,hur2022nonvolatile,luo2021design,zhou2022experimental}. To reliably detect the charge difference between two memory states, both designs require a large capacitor area, which limits the scalability of 1T-1C FeRAM. This issue is further exacerbated for capacitive sensing, given its small sense margin (typically less than 2$\times$), creating a trade-off between sense margin and read endurance in 1T-1C FeRAM.

Many of the challenges of 1T-1C FeRAM can be addressed with the 1T ferroelectric field effect transistor (FeFET) cell, which replaces the typical gate dielectric in a MOSFET with a thin ferroelectric film. 
In this cell, polarization can be configured to point towards the channel or the gate by applying an effective positive or negative gate pulses, respectively, which sets the device to the low-threshold voltage (LVT) state or high-threshold voltage (HVT) state, respectively, thus encoding a binary ‘1’ or ‘0’\cite{park2016ferroelectric,mulaosmanovic2021ferroelectric}. The intrinsic transistor structure and associated gain grant 1T-1FeFET cells excellent scalability, nondestructive read capability, and a large sense margin, effectively addressing the limitations of 1T-1C FeRAM. However, FeFETs introduce their own challenges, many of which stem from the ferroelectric/interlayer and interlayer/semiconductor interfaces\cite{gong2017study,yurchuk2016charge,zeng2019program}. First, due to the additional voltage drop across the semiconductor substrate, the write voltage for FeFETs is higher than that of MFM capacitor-based FeRAMs. Additionally, the depolarization field in the ferroelectric layer degrades retention, making it inferior to FeRAM but still acceptable for non-volatile memory (NVM) applications\cite{gong2016fe}. Over multiple write cycles, charge trapping and defect generation in the bulk ferroelectric and at interfaces gradually offset the polarization-induced threshold voltage (\textit{V}\textsubscript{TH}) shift\cite{yurchuk2016charge,gong2017study,zagni2023reliability}, leading to memory window (MW) degradation. This degradation significantly impacts FeFET reliability, severely limiting write endurance—for instance, Si-based FeFETs typically endure fewer than 10\textsuperscript{6} cycles\cite{ali2018high}. As a result, endurance remains a major hurdle to the widespread adoption of FeFET technology. 

\begin{figurehere}
   \centering
    \includegraphics[scale=0.1,width=\textwidth]{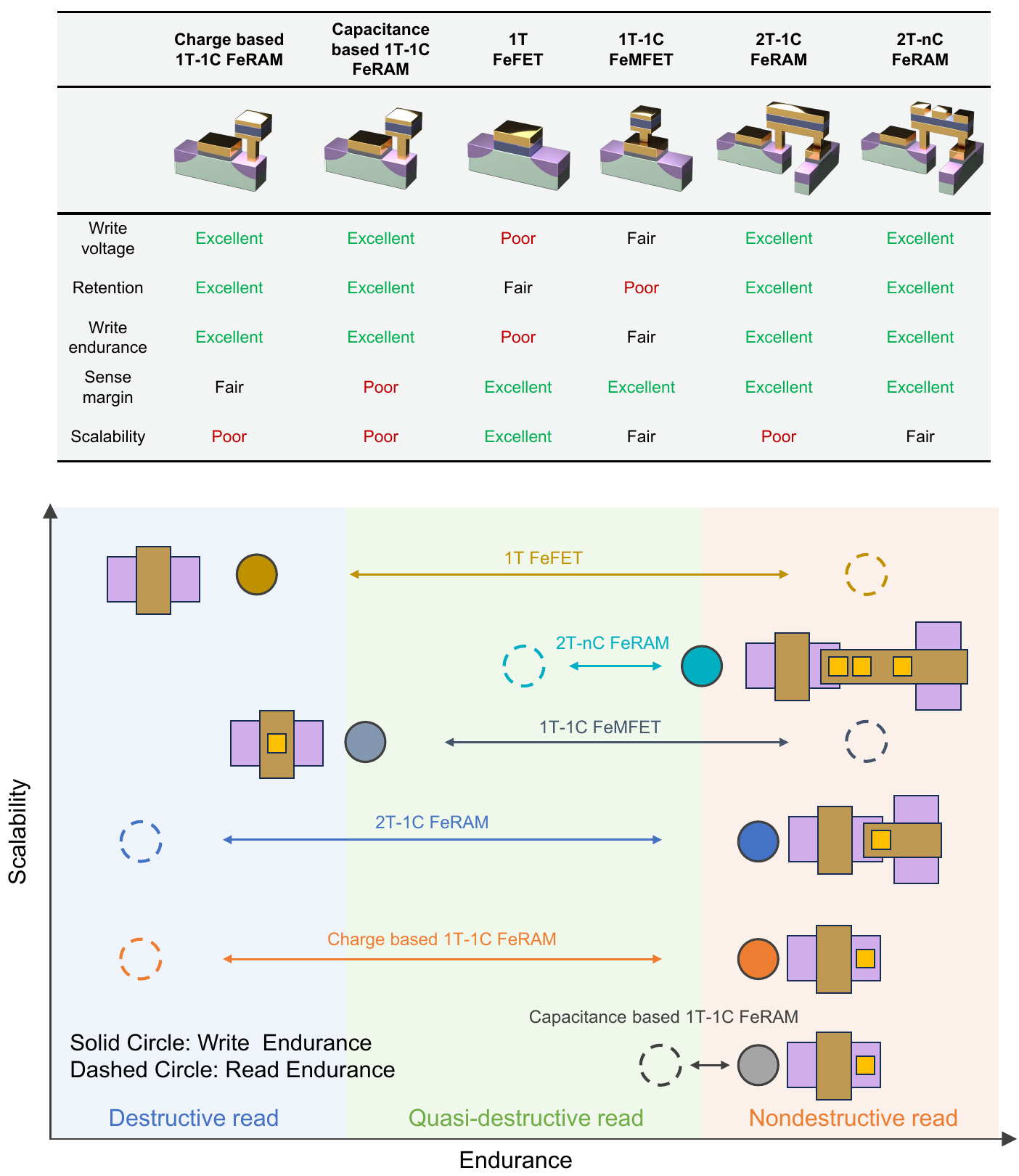}
    \captionsetup{parbox=none} 
    \caption{\textit{\textbf{Overview of ferroelectric memory cells.} Key performance and reliability metrics comparing various ferroelectric memory variants, including write voltage, retention, write endurance, read endurance, sense margin and scalability for charge based 1T-1C FeRAM, capacitance based 1T-1C FeRAM, 1T FeFET, 1T-1C FeMFET, 2T-1C FeRAM, and 2T-nC FeRAM.}}
    \label{fig:Ferroelectric Memory Classification}
\end{figurehere}

Identifying the primary source of reliability issues in 1T-FeFET as the ferroelectric/interlayer and interlayer/semiconductor interfaces, a modified device architecture has been developed. This variant incorporates a metal-ferroelectric-metal (MFM) capacitor atop the gate of a MOSFET, forming a metal-ferroelectric-metal-insulator-semiconductor (MFMIS) stack, commonly referred to as FeMFET\cite{ni2018soc,kazemi2020hybrid}. By isolating the ferroelectric layer from the semiconductor channel, this design effectively mitigates many of the associated challenges. 
In a 1T-1C FeMFET, the ferroelectric capacitor can be implemented either in the back-end-of-line (BEOL) or integrated with the Si MOSFET in the front-end-of-line (FEOL) process, and it is electrically connected to the gate of the MOSFET. This separation provides significant design flexibility, particularly in optimizing the area ratio between the ferroelectric capacitor and the MOSFET gate — a key factor in controlling the memory window and ensuring reliable switching behavior\cite{lee2021ferroelectric,wang2024comprehensive,wang2024unveiling,wang2023first,ali2022impact,seidel2022memory}.
Reducing the ferroelectric capacitor size relative to the MOSFET area maximizes the voltage drop across the ferroelectric layer, facilitating efficient polarization switching at lower write voltages. Additionally, the 1T-1C FeMFET improves endurance\cite{hu2019split}, as the isolated MFM capacitor absorbs most of the write voltage, minimizing stress on the transistor. This reduces charge injection, suppresses defect generation, and mitigates cycle-dependent degradation, ultimately extending device lifetime.
However, the introduction of a floating metal in the FeMFET can degrade retention due to leakage from the floating node between the MOSFET and the MFM capacitor, leading to charge loss and instability over time\cite{sun2021temperature,yoon2020improvement}. Moreover, if an optimal area ratio smaller than 1 is adopted, even if the MFM capacitor is defined with lithography limits, the MOSFET area will be large, thus limiting its scalability.

Based on the observation that MFM capacitor offers excellent write performance and reliability, while the transistor memory supports charge amplification through its gain that offers a large sense margin and scalability, hybrid cells, such as 2T-1C FeRAM\cite{hur2021technology,sun2025back,slesazeck2019uniting,luo2021technology,wu2023two,10288349} and 2T-nC FeRAM\cite{slesazeck20192tnc,xiao2023quasi,deng2023comparative,deng2024first}, have been developed to harness the strengths of both capacitor and transistor memories are developed. In these designs, two transistors are adopted, one as write access transistor like that in 1T-1C FeRAM, responsible for the write access and programming the polarization and the other one is a read transistor, responsible for providing gain during the read operation.
Since hybrid cells share the same write path as 1T-1C FeRAM and store information in the MFM capacitor, they maintain similar write performance and reliability, including low write voltage, high write endurance, and excellent retention. During read operations, the write transistor is disabled, and a read pulse is applied to the MFM top electrode while sensing the read transistor’s channel current.
In a typical 2T-1C configuration, polarization is destructively switched and amplified through the read transistor\cite{slesazeck2019uniting,hur2021technology}. This enhances the sense margin and capacitor scalability, though a write-back operation remains necessary, as in 1T-1C FeRAM. Building on the 2T-1C FeRAM concept, it was later recognized that multiple MFM capacitors could share the two transistors, reducing transistor overhead. Additionally, complete polarization switching is unnecessary, as turning on a typical transistor requires a charge density of only $\sim$0.2 $\mu$C/cm\textsuperscript{2}, whereas HfO\textsubscript{2} typically has a 2\textit{P}\textsubscript{r} of 50 $\mu$C/cm\textsuperscript{2}.
This large gap leads to the proposal of the 2T-nC FeRAM cell, which offers high density, scalability, and quasi-nondestructive readout capability (QNRO)\cite{xiao2023quasi, yoon2001memory, kato20050, horita2008nondestructive}. The large charge gap between the capacitor’s charge capacity and the required switching charge enables multiple read cycles before a write-back is needed. During read operations, only a small fraction of the polarization is switched on top of the difference of the nonlinear small-signal capacitance—allowing it to be sensed by the read transistor without fully disrupting the polarization. Additionally, some of the switched domains during read could flip back after the read pulse, further extending the read endurance. Since the write process remains the same as in 2T-1C FeRAM, the write voltage, retention, and endurance are also comparable. One unique challenge that 2T-nC brings is the disturb management during the device operation as multiple capacitors share the same signals, as will be discussed in the 3D array design\cite{nishihara2002quasi}. Although it is not explicitly shown here, there are also on the 1T-nC FeRAM cell\cite{haratipour2022hafnia,luo2024endurance,lim20253d}, where multiple ferroelectric capacitors share one access transistor without read transistor compare to 2T-nC FeRAM. However, it also faces the challenges from the disturb management, similar to 2T-nC FeRAM.

\section*{\textcolor{sared}{\large Polarization Sensing Mechanism}}

\begin{figurehere}
   \centering
    \includegraphics[scale=0.1,width=\textwidth]{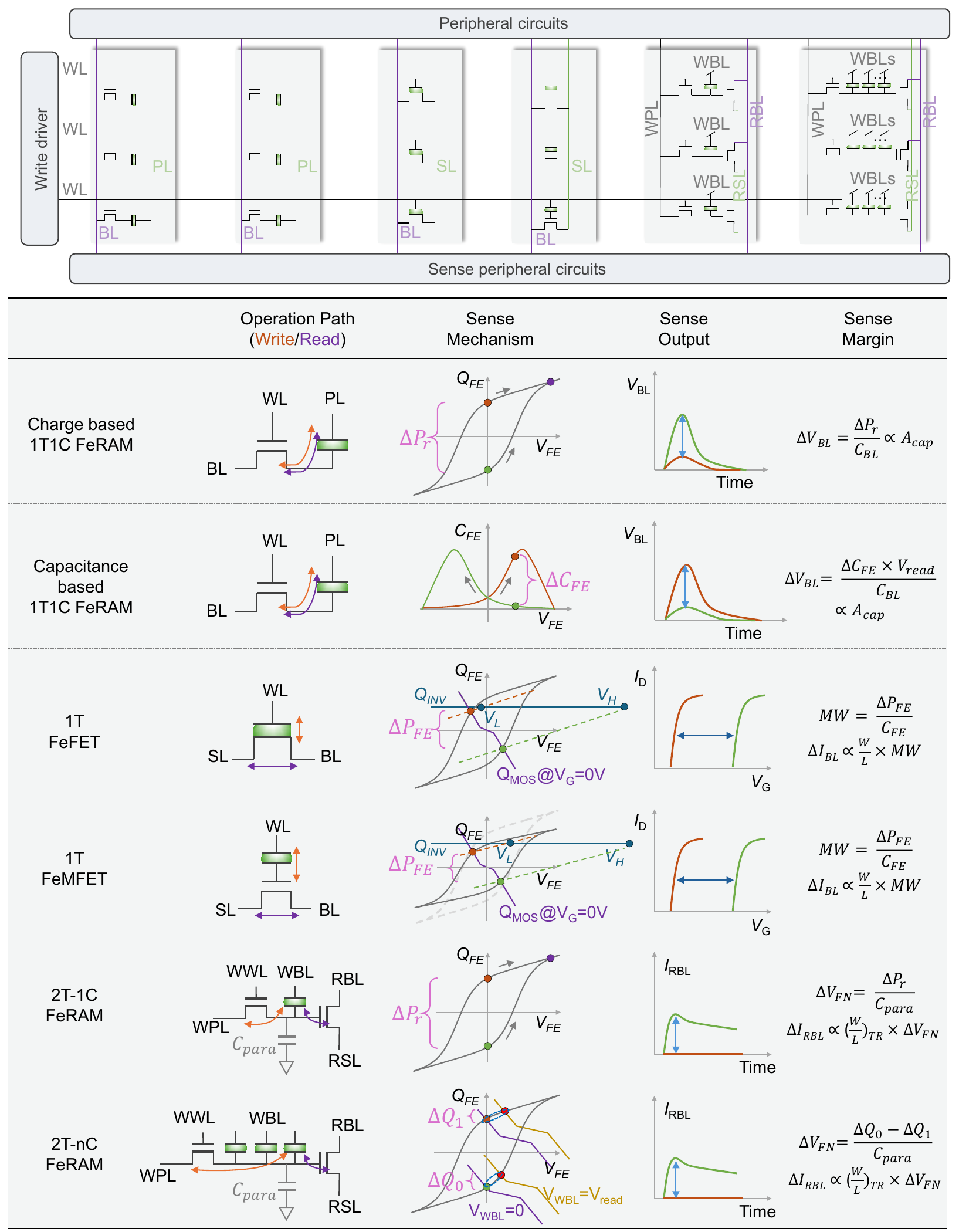}
    \captionsetup{parbox=none} 
    \caption{\textit{\textbf{Summary of sensing mechanisms of different ferroelectric memory cells.} The charge based and capacitance based 1T-1C FeRAM share the same write and read path, while charge based sensing directly read the polarization current when a high voltage is applied at BL. The capacitance based 1T-1C FeRAM utilizes the different dielectric constants at different polarization states to distinguish the data, enabling nondestructive read. For these two FeRAM designs, the sense margins are both proportional to area of the capacitor. The write paths for 1T FeFET and 1T-1C FeMFET are from the gate to the source/drain, while the read path is from source to drain. The sense margin is determined by the W/L ratio and also can be understand from the loadline of MOSCAP and polarization-voltage curves of MFM capacitor. The write paths of 2T-1C and 2T-nC FeRAM are both from WPL to WBLs, which is controlled by the write transistor. And the read paths are from WBLs to RBL.}}
    \label{fig:Sensing mechanism}
\end{figurehere}

After the overview of the different types of ferroelectric memory, the detailed sensing mechanisms are discussed. The sense mechanisms and correlated sense margin for the ferroelectric memory designs are shown in Fig. \ref{fig:Sensing mechanism}. To clearly demonstrate the operation, one column of each type of ferroelectric memory cell are shown. 
In 1T-1C FeRAM cell, the gate and drain of the access transistor are connected to wordline (WL) and bitline (BL), respectively. The two terminals of MFM capacitor are connected to the source of access transistor and the plateline (PL). To write this cell, a positive voltage is applied at BL or PL for data "1" or data "0", while access transistor is turned on. For charge-based 1T-1C FeRAM sensing, a sufficiently high read voltage is applied to the PL to fully switch the ferroelectric capacitor from the upward to the downward polarization state. The resulting switching current directly reflects the stored state and is sensed through the BL voltage. If the stored polarization state is upward, the sensed current is high, which will raise the BL voltage; if the polarization state is downward, the sensed current is low and the resulting BL voltage remains low. However, since the upward polarization state is destroyed after sensing, this approach suffers from poor read endurance, necessitating a write-back operation, which requires sufficient endurance for the cell. The sense margin of the cell can be expressed as\cite{takashima2011overview,oh2024design}:
\begin{equation}
    \Delta V_{BL} = \frac{\Delta P_r}{C_{BL}} \propto A_{cap}
\end{equation}
where the $\Delta P_r$, \textit{C}\textsubscript{BL}, and \textit{A}\textsubscript{cap} are the switched polarization, the BL capacitance, and the capacitor area, respectively. With aggressive scaling, to keep sufficient sense margin, a large effective capacitor area at a small area footprint is required, thus requiring 3D capacitor structure and posing integration challenges.
As explained before, capacitance-based 1T-1C FeRAM mitigates the destructive read issue. In this approach, a small read voltage is applied to the BL to sense the capacitance difference without switching the polarization state of the ferroelectric capacitor\cite{kato20050,mukherjee2023capacitive,9720508,luo2020non,xiang2024compact,lou2024super,zhou2020metal,10873459}. The output is reflected also on the BL voltage. The sense margin is expressed as\cite{mukherjee2023pulse}:
\begin{equation}
    \Delta V_{BL} = \frac{\Delta C_{FE} \times V_{read}}{C_{BL}} \propto A_{cap}
\end{equation}
where the $\Delta C_{FE}$ is the small signal capacitance difference between the two states and $V_{read}$ is the read voltage. 
In this case, the margin is also proportional to the capacitor area. However, given its small capacitance ratio (typically less than 2$\times$), it will require an even larger capacitor area than the conventional 1T-1C FeRAM. Although it is possible to improve the capacitance ratio through stack engineering\cite{yu2024ferroelectric,mukherjee2023pulse}, especially introducing nonlinear semiconductor capacitance\cite{zhou2020metal,10873459}, the effectiveness is compromised if the semiconductor layer thickness is constrained for high density integration.

For the transistor based memory, such as 1T-FeFET and 1T-1C FeMFET, their operations are similar. In both cells, WL, BL, and source line (SL) are connected to the gate, drain and source terminals, respectively. To program the memory, voltages on WL and necessary inhibition bias schemes applied to BL and SL are exerted to set the device to the LVT or HVT while keeping unselected cells intact\cite{ni2018write,jiang2022feasibility,xiao2022write,kim2024exploring,venkatesan2024disturb}. After polarization is programmed, sensing of the FeFET is typically conducted in the current domain where a proper read gate voltage between the LVT and HVT is applied to WL, which could induce different drain currents for the LVT and HVT states. 
The sensing operation of 1T-FeFET can be understood by examining the loadline of the semiconductor channel and the ferroelectric polarization hysteresis loops\cite{toprasertpong2022memory,yoo2023analytical,zheng2024beol,qin2024understanding}. Since the semiconductor channel cannot provide screening charge for all the polarization, the ferroelectric will likely operate in minor hysteresis loops. After programming, and when the WL voltage returns to ground, there are two intersection points between the channel charge and the hysteresis loop, corresponding to the LVT and HVT states, as shown in Fig.\ref{fig:Sensing mechanism}. It is assumed that during sensing, the polarization remains fixed enabling nondestructive read such that the linear ferroelectric capacitance supports the read operation. In this way, the memory window and the sense margin can be expressed as:
\begin{align}
        MW &= \frac{\Delta P_{FE}}{C_{FE}} \label{eq:3}\\
    \Delta I_{BL} &\propto \frac{W}{L}\times MW  \label{eq:4}
\end{align}
where the MW is the voltage separation between the LVT and HVT states and the current sense margin is estimated based on the assumption that HVT yields negligible current while the LVT state works at the linear region during sensing. 
Because this margin is not proportional to the area of the transistor, but the W/L ratio, the scalability is better than that of other ferroelectric memory devices.

The cell structure, write scheme, and sense method for 1T-1C FeMFET are almost the same as that of 1T-FeFET. But the flexibility of tuning the area ratio of ferroelectric capacitor and MOSFET offers the opportunity to decrease the write voltage and improve the write reliability\cite{ni2018soc,dahlberg2025memory,hwang2024effect,wang2024comprehensive}. In typical cases, a smaller area ratio is favored to increase the ferroelectric voltage drop. Therefore, from the loadline analysis shown in Fig.\ref{fig:Sensing mechanism}, the smaller ferroelectric area leads to close-to-saturation-loop operation and it changes the intersection points for the two memory states compared with the 1T-FeFET. Other than this, the Eq.(\ref{eq:3}) and (\ref{eq:4}) are still applicable. Therefore, like that in FeFET, the current sense margin is proportional to the W/L ratio, leading better scalability than FeRAM. However, if a smaller ferroelectric capacitor is used, the transistor size cannot be minimum, leading to a slightly degraded scalability than 1T-FeFET. 

In the 2T-1C and 2T-nC FeRAM, the gate and drain of the write transistor are connected to write word line (WWL) and write plate line (WPL), respectively. The source and drain of read transistor are connected to read bit line (RBL) and read source line (RSL), and its gate is connected to the common terminal of the MFM capacitors along with the source of the write transistor. The other terminal of MFM capacitors are connected to separate word bit line (WBL). The 2T-1C and 2T-nC both utilize the write scheme of 1T-1C FeRAM, i.e., by enabling the WWL and bias WBL or WPL to write the polarization. To read 2T-1C FeRAM cell, write transistor is turned off, while the WBL is applied at a fixed read voltage and RBL is activated in order to sense the RBL current. In the original designs, full polarization switching is assumed, which can raise the internal floating node potential $V_{FN}$\cite{wu2023two,10288349} and the difference of the node voltages $\Delta V_{FN}$ and the current margin are:
\begin{align}
        \Delta V_{FN} &= \frac{\Delta P_{r}}{C_{para}} \label{eq:5}\\
    \Delta I_{RBL} &\propto (\frac{W}{L})_{TR}\times \Delta V_{FN}  \label{eq:6}
\end{align}
where $C_{para}$ and (W/L)\textsubscript{TR} are the parasitic capacitance on the floating node and the read transistor W/L, respectively. In this case, if the MFM capacitor and the MOSFET scales, the $C_{para}$ scales at an almost the same rate, making the sense margin almost independent of the area to the first order, thus improving the scalability over the 1T-1C FeRAM. 
As for 2T-nC FeRAM, QNRO is possible by adopting an appropriately chosen voltage at WBL to switch a tiny portion but enough polarization to sense the RBL current difference of the two memory states. The loadline analysis shown in Fig.\ref{fig:Sensing mechanism} illustrates the sensing operation. Under the applied WBL voltages, state '0' switches $\Delta Q_0$, which is higher than state '1' $\Delta Q_1$, thereby causing a difference in the floating node voltage $\Delta V_{FN}$:
\begin{equation}
    \Delta V_{FN} = \frac{\Delta Q_{0} - \Delta Q_{1}}{C_{para}} \label{eq:7}
\end{equation}
and the resulting RBL current sense margin will be the same as 2T-1C FeRAM case shown in Eq.(\ref{eq:6}). In this case, due to more than one MFM capacitors connecting with the floating node, the $C_{para}$ could be higher than the 2T-1C FeRAM case. But with a fixed number of capacitors, the margin, to the first order, remains fixed with the geometry scaling, thus showing better scalability than 1T-1C FeRAM. One more thing worth noting is that after the read pulse is applied, some of the switched polarization can potentially flip back, extending the number of read endurance cycles. This is illustrated by the small minor hysteresis loops indicated in Fig.\ref{fig:Sensing mechanism} loadline analysis. If the read operation is constrained on this loop, then the read operation becomes almost free from disturb. However, in reality, repetitive read will induce accumulative switching and eventually operate beyond the minor loop and cause read degradation.     


\section*{\textcolor{sared}{\large 3D Integration of Ferroelectric Memory}}
The analysis of polarization sensing mechanisms reveals their critical role in determining the cell's sense margin, design, and scalability. In the following, dense 3D integration methods for various types of ferroelectric memories are summarized and compared.
Typically, two primary 3D integration methods are employed: parallel stacking and sequential stacking. In parallel stacking (Fig.\ref{fig:Parallel Stacking}), fabrication follows a top-down approach with a limited number of lithography steps, remaining largely independent of the number of stacked layers. This approach becomes increasingly cost-effective as the number of stacked layers grows, which is essential for achieving high-density storage. The process involves a combination of etching and deposition steps, enabling efficient manufacturing.
In contrast, sequential stacking (Fig.\ref{fig:Sequential Stacking}) constructs 3D arrays layer by layer, offering greater flexibility in material selection and device integration. However, lithography and processing complexity scale nearly linearly with the number of stacked layers, resulting in higher fabrication complexity and longer processing times, particularly when many layers are involved. Additionally, cells in different layers undergo varying thermal processes, leading to performance variations across layers. Consequently, sequential stacking is less suitable for stand-alone storage but is more commonly adopted for embedded applications. In the following sections, both parallel and sequential stacked 3D ferroelectric memories are presented. As summarized in Fig.\ref{fig:Sensing mechanism}, the constraints imposed by polarization sensing mechanisms dictate the structural design of 3D arrays, necessitating adherence to the limitations posed by their respective information sensing methods.

\subsection*{\normalsize Parallel Stacking 3D Integration}
\begin{figurehere}
   \centering
    \includegraphics[scale=0.1,width=\textwidth]{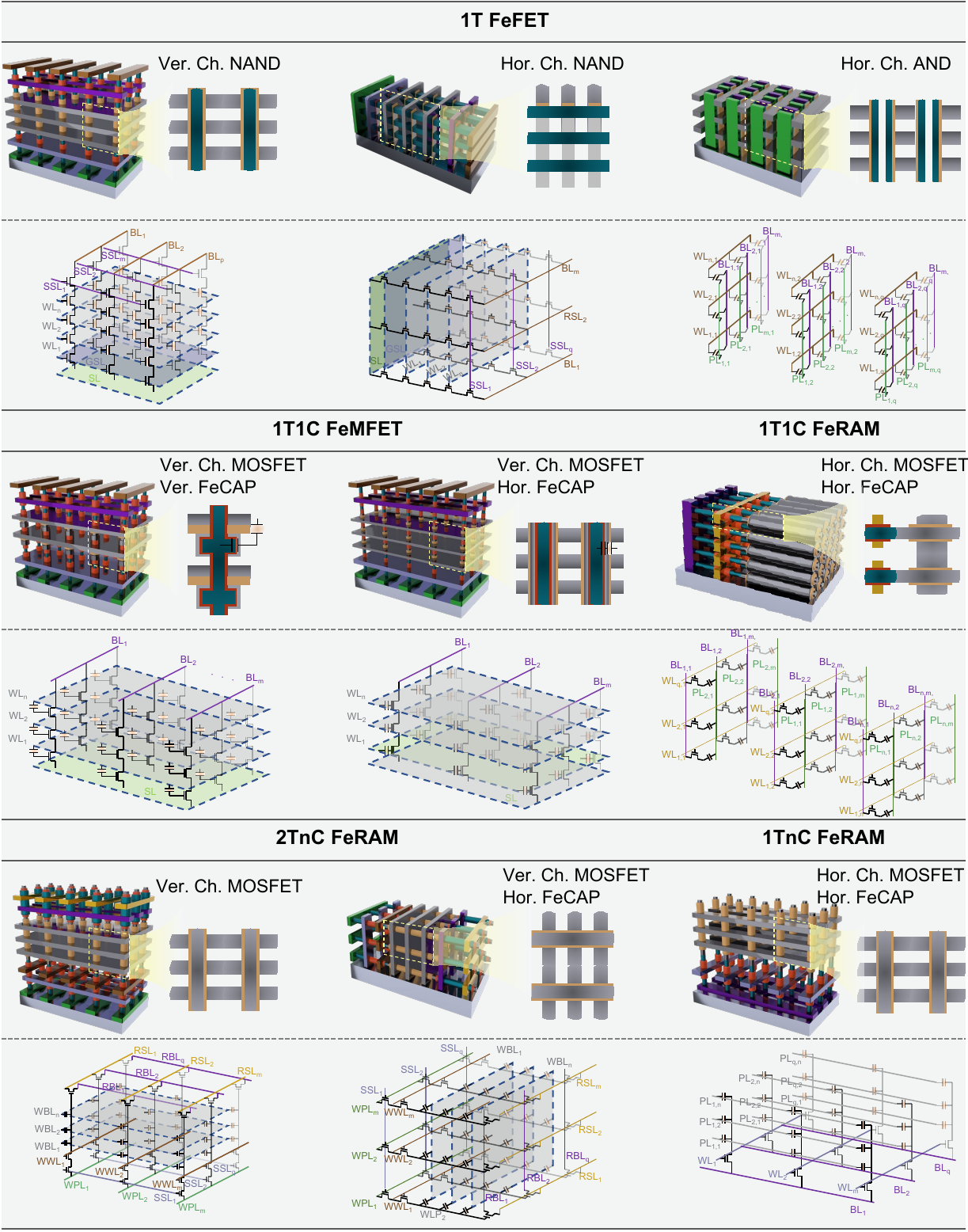}
    \captionsetup{parbox=none} 
    \caption{\textit{\textbf{3D parallel stacking of ferroelectric memories.} Parallel 3D integration has the advantages of bit-cost scalability due to the parallel processing that the processing costs does not increase linearly with the number of layers and also uniform thermal treatment for different layers. 3D Vertical NAND, horizontal NAND and also vertical AND designs are available for 1T FeFET; Veritcal NAND structure can also be adopted in 1T-1C FeMFET, and the ferroelectric capacitor can run vertically and horizontally.  3D DRAM integration strategy can also be applied for 1T-1C FeRAM. 2T-nC can also adopt the vertical or horizontal NAND like 3D design, where multiple ferroelectric capacitors can be integrated in NAND-like structure. When avoiding the read transistor, the 2T-nC becomes 1T-nC, which is also actively pursued.}}
    \label{fig:Parallel Stacking}
\end{figurehere}

For parallel stacking 3D integrated FeFET, three exemplary designs are shown: Fe-NAND with vertical channel transistors (VC-FeNAND)\cite{florent2018vertical,banerjee2021first,yoon2022highly,yoon2023qlc}, Fe-NAND with horizontal channel transistors (HC-FeNAND) and Fe-AND with horizontal transistors\cite{feng2024first}. Due to its excellent scalability, small transistors can be adopted for all the designs. The NAND array is widely used in high density storage\cite{tanaka2007bit}, such as the vertical NAND flash due to its dense cell and small overhead for the contact routing (one BL per string). The AND array, on the other hand, requires routing for both the source and drain, increasing the overhead. But it offers fast read operation as, unlike the NAND array, the cell can be read out directly without the need of passing through the whole string. In 3D Fe-NAND, FeFETs are connected in a vertical NAND string, where each FeFET shares a common semiconductor channel (typically poly-Si), with the source of one FeFET linked to the drain of the next in series. Due to its high polarization switching efficiency, FeFET has received significant attention as a replacement for the flash transistor, which relies on the highly inefficient tunneling process for programming. In such an array, word lines (WLs) run horizontally across all strings in a block, controlling the ferroelectric gates of multiple FeFETs. The BL connects to the top of the vertical FeFET string through one or several top select transistors, whose gate is connected to string select line (SSL) running horizontally and perpendicular to the BLs. The source line plane (SL) is at the bottom, which is connected to the bottom select transistor, whose gate is connected to ground select line (GSL). The difference between VC-FeNAND and HC-FeNAND lies in the channel direction of the FeFET, while the connections remain the same. The horizontal design is advantageous in that the string current is independent of the number of stacked layers unlike the vertical channel where string current decreases with the number of stacked layers\cite{yeh2016z}. Also, the density could be further increased by employing the local block interconnect architecture, which connects several blocks or string in advance so that the they can share the staircases, consequently increasing density\cite{oda2024superior}. But the manufacturing simplicity and high density make the VC-FeNAND a preferred architecture. Additionally, the channel length is defined by the thickness of the WL, rather than via sophisticated lithography as in the HC-FeNAND design, making scaling more straightforward. To ensure selective programming, an inhibition bias scheme needs to be employed to prevent unintended cells from being programmed. This is typically achieved by grounding the target string while leaving the other unselected strings floating such that the channel potential is boosted through coupling to WLs to prevent unintentional programming\cite{kim2021cmos,kim2024exploring}. To read a single cell, pass voltages are applied to all word lines (WLs) except for the target WL, which is driven by the read voltage to distinguish the HVT and LVT states of the target FeFET. The resulting bit line (BL) current of the string is then sensed. The high switching efficiency of FeFETs also introduces a new challenge—pass disturb. During string operation, a high pass voltage is required, making the HVT state particularly vulnerable to disturbance \cite{zhao2024paving}. To mitigate this issue, approaches such as gate stack optimization \cite{kim2024exploring} and the use of double-gate FeFETs—where a dedicated pass gate with a non-ferroelectric gate oxide is employed—have been proposed \cite{zhao2024paving}. Nevertheless, further investigation is essential to enable the practical implementation of 3D NAND FeFET storage.
Another unique challenge in FeFET NAND is the erase-verify operation. Unlike flash memory, the erase process in FeFET NAND sets all transistors in a string to the HVT state. Any FeFET that remains in the LVT state fails to erase and must be identified to allow repeated erase cycles until all FeFETs in the block are successfully erased. However, detecting a FeFET in the LVT state in a string of serially connected FeFETs in the HVT state is nontrivial. This issue can be addressed by leveraging the fact that HVT/LVT states block/pass electron conduction but pass/block hole conduction, respectively. Thus, while normal operations rely on electron conduction, erase-verify can utilize hole conduction. Two methods have been proposed: introducing a p+ doping region to supply holes \cite{maeda2022semiconductor}, or generating holes through the gate-induced drain leakage (GIDL) effect \cite{kumar2025erase}. However, the validity of these techniques still needs to be studied.

Typically, the poly-Si is employed as the channel material in 3D NAND, so a high process temperature is needed, which means the ferroelectric layer will go through this thermal process as well. Thus, the thermal stability of ferroelectric materials is another crucial consideration in sequential stacking. Consequently, materials like Si- or Al-doped HfO\textsubscript{2}, which retain robust ferroelectric properties at relatively high temperatures\cite{yang2023ferroelectricity} are often preferred for such applications in poly-Si channel NAND\cite{banerjee2021first,florent2017first}. Recently, the metal oxide semiconductor FeFET gains lots of interest due to the lower deposition temperature of the metal-oxide channel, making it also a candidate for 3D NAND FeFET application. However, a key challenge in applying FeFETs for 3D NAND storage is their limited memory window at constrained gate stack thicknesses (necessary for small string diameters) \cite{qin2024clarifying} for both poly-Si and metal oxide semiconductor channels. This limitation hinders the ability to store multiple bits per cell, reducing logical density and making FeFETs less competitive than conventional NAND flash. However, recent advancements in gate-side injection techniques \cite{rajwade2011ferroelectric} aim to overcome the trade-off between memory window and gate stack thickness, making FeFETs viable for vertical NAND storage \cite{myeong2024strategies,lim2023comprehensive,zhao2024large,qin2024clarifying,joh2024oxide,yoo2024highly}.
One common approach is to insert a trapping layer and an additional tunneling layer between the ferroelectric material and the gate metal. This structure, combined with gate-side injection, allows the incremental step pulse programming (ISPP) slope to reach and even exceed the theoretical limit of 1 in conventional flash transistors\cite{yoo2024highly,kim2024unveiling}, enabling low-power operation. However, this improvement also introduces retention issues likely caused by the detrapping of charges injected from gate side and polarization loss facilitated by depolarization field, which remains an open challenge to be addressed. In addition, the impact of scaling on the device variability still needs to be examined, especially considering the impact of limited number of domains in ferroelectric in highly scaled dimensions. Another challenge for NAND FeFET is the erase operation. Due to the GAA structure in 3D NAND FeFET, no holes are available to support the erase operation. The issue is exacerbated for most metal oxide semiconductors due to its natural n-type doping and wide bandgap. To address this issue, the gate-induced drain leakage (GIDL) assisted erase operation is employed in 3D NAND with poly-Si channel \cite{komori2008disturbless,malavena2018investigation,ryu2023selective}. Given that GIDL is unlikely in wide bandgap oxide channel, alternative approaches are being explored. For example, Samsung reported that insertion of a thin oxygen-deficient layer between the channel and ferroelectric can provide sufficient positive depletion charges during erase operation\cite{yoo2024highly}, addressing the challenge of hole deficiency in metal-oxide semiconductors at the cost of increased leakage current. By adopting both gate-side injection and the intermediate oxygen-deficient layer in 3D FeFET NAND, the record-largest MW (17.8 V) based on the IGZO channel FeFET is reported\cite{yoo2024highly}. 


In contrast to NAND-based ferroelectric memory, which uses series-connected cells, the AND structure arranges memory cells with parallel connection\cite{feng2024first,lue20183d,lue2018novel,wei2022analog,lue2021investigation}. The WLs runs horizontally, while SL and BL run vertically. FeFETs in each string share one BL and SL. To independently program cells in this AND array, target WL is applied at program voltage and target BL and SL are grounded, while unselected BLs and SLs are applied inhibition biases to prevent these cells from programming. To sense the target cell, the target WL, SL, BL are biased at V\textsubscript{read}, GND, and $V_D$, respectively. Unselected WLs are biased at GND so that the current of the target BL is totally attributed to the select FeFET. This architecture offers faster access times because it is not necessary to turn on the other FeFETs in the string where target cell is located. As can be seen in the array structure, in AND array, it is necessary to connect each BL and SL, doubling the routing resources compared with the NAND array, where contact to BL needs to be routed. Therefore, AND array is more attractive for applications requiring fast read speed, such as inference applications for neural networks. Recently, the 3D NAND process compatible 3D AND-flash design is proposed\cite{lue20203d,wei2022analog,lue2021investigation}, which enables fast read time, high sensing current, and high integration density potential for future storage-class memory application.

The 1T-1C FeMFET, sharing a similar structure as 1T FeFET, can also be arranged in NAND or AND array. Here, for brevity, the vertical channel NAND array is illustrated and only highlighting the aspect that differs from 1T FeFET. 
In this configuration, the MFM capacitor, which plays a critical role in the device’s operation, and can have a different area than the semiconductor. This complicates the integration. To achieve this flexibility, two designs have been proposed\cite{woo2023design,kim2022high}, as shown in Fig.\ref{fig:Parallel Stacking}. The capacitor can be integrated into the design in a manner similar to the gate formation of a vertical channel MOSFET\cite{woo2023design}, allowing for a more compact and efficient structure. An alternative design involves a fully vertical arrangement\cite{kim2022high}, where the capacitor is aligned with the channel to enhance device performance. Both of these 3D integration methods share a similar framework and operational principles to those used in VC-FeNAND technology, which has been shown to deliver high-density storage solutions.
 
 The vertical stacking 3D 1T-1C FeRAM architecture utilizes monolithic stacking to achieve high memory density and performance, similar to the 3D DRAM\cite{huang20233d,10185290}. For each layer, a horizontal gate-all-around (GAA) access transistor and horizontal MFM or metal-ferroelectric-semiconductor (MFS) capacitor can be adopted. Given the nonvolatility of ferroelectric, there is no need of single-crystalline silicon channel for ultra-low leakage, thus opening the space for low-cost channel solutions, such as poly-Si and amorphous oxide semiconductor. With the same array core, there are two variants of the WL/BL routing: one is the vertical BL and horizontal WL and the other is the horizontal BL and vertical WL\cite{10185290,wu2025signal}. As the former routing scheme exhibits a vertical layer independent sensing delay, it is typically preferred and illustrated in Fig.\ref{fig:Parallel Stacking}\cite{10185290}. In this structure, WLs are horizontally placed to control the gate of the access transistors, BLs run vertically to connect the source/drain of transistors for data access, and horizontal PLs are used to apply voltage across the ferroelectric capacitors to induce polarization switching. During a write operation, the target WL is activated to turn on the access transistor of the target cell. And a voltage difference is applied between the target BL and PL to polarize the target ferroelectric capacitor, storing data as a stable remnant polarization state. In a read operation, the target WL activates the target access transistor, and the stored polarization state is sensed as the BL voltage. Compared with the other variant with horizontal BL and vertical WL, the vertical BL 3D 1T-1C FeRAM exhibits lower BL capacitance and reduced BL-to-BL coupling, resulting in a larger sense margin as the number of stacking layers increases. In contrast, horizontal BL architectures require an even tighter staircase pitch than that in 3D NAND to maintain sufficient sense margin, posing greater challenges for the integration process. Additionally, the contact heights of different layers in horizontal BL design are different, which leads to the sense margin for top tiers being larger than that for bottom tiers. Beside this GAA access transistor based 3D DRAM, channel-all-around access (CAA) transistor with Indium-Gallium-Zinc-oxide (IGZO) channel 3D DRAM design is also reported with bit cost-scalable process and long retention time\cite{ai2024first}, which could also a potential design for 3D 1T-1C FeRAM.
 
It is very challenging, if not impossible, to integrate 2T-1C FeRAM via parallel integration processes as each cell requires separated write and read transistors. Luckily, parallel stacking is feasible for 2T-nC FeRAM and there are two typical 3D integration architectures with horizontal string or vertical string. They are simply 90$^\circ$-flipped structures of each other. Each string corresponds to a 2T-nC cell. In vertical 2T-nC FeRAM, the number of layers corresponds to the number of MFM capacitors in one cell. As the number of layers stacked increases, the electrical performance of the cell may degrade as more parasitics on the floating node are introduced. The array resembles the vertical NAND array that transistors are on the top and bottom and the middle section sits the memory cells. However, unlike the NAND array, the core of the string is metal, not semiconductor channel, making the cells in a string in parallel, not in series. 
There are one write transistor and one or more string select transistors at the bottom of each string, which selectively pass the WPLs to write to the ferroelectric capacitors. And there is also one read or gain transistor at the top of each string for amplifying the charge current to read the polarization state. To write the target cell and avoid the unintended programming to unselected cells, inhibition bias schemes need to be applied such that the target cells experience the full write voltage and unselected cells experience a half or a third of the write voltage\cite{nishihara2002quasi,deng2023comparative,deng2024first}.This is achieved by applying appropriate biases on the WBLs and charging the internal nodes to different voltages through the WWL, SSL, and WPL combination. As for read operation, all WWL and SSL are grounded to disable the write transistors and target WBL, RBL, and RSL are activated to sense out the current. In the horizontal 2T-nC FeRAM, one benefit is that the number of MFM capacitors in one cell, i.e., \textit{n}, is independent of the number of layers vertically stacked, thus preserving the electrical integrity with scaling. Additionally, the process integration of this architecture is similar to the monolithic 3D 1T-1C FeRAM, except that multiple capacitors are integrated. On top of that, the read transistor integration is challenging, which requires channel-all-around transistors. Compared with horizontal string configuration, the components in vertical string configuration are easier to integrate using GAA access transistor, providing better gate control of access transistor in same footprint, this also leads to a larger sense margin. Furthermore, benefiting from the mature 3D NAND process, the vertical string integration approach is the preferred choice.
 
Beyond the 2T-nC FeRAM, there are also interests in the 1T-nC FeRAM structure, which has multiple MFM capacitors and only one single write transistor in a cell\cite{lim20253d,florent2017first,van20173d}. Therefore, it is similar to the 1T-1C FeRAM that memory information is read out by directly sensing the polarization charge. A major challenge in this architecture lies in scalability, i.e., the smallest cell size that is required for correct sensing and the ultimate memory density.
As shown in Fig.\ref{fig:Parallel Stacking}, the differences between the 1T-nC FeRAM and 1T-1C FeRAM is that the write transistor is shared among multiple capacitors. Therefore, similar inhibition bias schemes as in the 2T-nC array need to be applied and the capacitors could be susceptible to disturb issues, at the cost of high density. Recently, disturb management in ferroelectric memories has gained increasing attention. One proposed solution involves integrating a selector with each ferroelectric capacitor to suppress the disturb voltage during inhibition bias while preserving the write voltage during write operations \cite{deng2025vertical}. Among the selector options, the two-terminal metal–semiconductor–metal (MSM) back-to-back Schottky diode has demonstrated effectiveness, though it requires higher write voltages compared to standalone ferroelectric capacitors. Further research is needed to achieve a more comprehensive and optimized solution. 
 
\subsection*{\normalsize Sequential Stacking 3D Integration}
Sequential stacking, through layer-by-layer stacking of the 2D planar designs, are also gaining attention as high density back-end-of-line (BEOL) embedded memory\cite{datta2024amorphous,datta2019back}. The complexity added compared with the planar version is the signal routing and the thermal challenges to the bottom tiers when integrating top ones. The complexity in signal routing arises from the necessary combination of vertical vias, horizontal interconnects, and shared circuitry for efficient connectivity across stacked layers such that the alignment precision and the parasitic capacitance and resistance in interconnects should be carefully considered. Additional resources dedicated for interconnects also degrade the array area efficiency, compared with the parallel 3D counterparts. 
The thermal challenge requires high thermal stability of the channel and ferroelectric materials, and reliable passivation layers between stacking layers, are essential to protect the bottom tiers from subsequent processing steps. Additionally, to enable BEOL integration, the deposition and annealing processes for the channel and ferroelectric layers must adhere to a strict thermal budget to avoid disrupting the bottom-tiers memories and FEOL circuitry. 

In sequential designs, various channel materials can be employed, including metal-oxide semiconductors, 2D materials, and polycrystalline silicon (poly-Si). Among them, metal-oxide semiconductors especially doped In\textsubscript{2}O\textsubscript{3} materials have garnered considerable interest due to their low processing temperatures and satisfactory mobilities even as channel thickness scales down, while poly-Si channel requires a relatively higher thermal budget for crystallization and suffers from significant mobility degradation as thickness decreases. Also, channel stability remains a critical concern, encompassing issues such as bias stress instability and environmental sensitivity. As metal-oxide semiconductor channel materials typically has larger trap states density compared to poly-Si, bias stress instability arises primarily from oxygen vacancy generation and hydrogen ion migration under prolonged electrical stress\cite{lee2022analysis,choi2023mechanism,yang2021total,wu2022characterizing,kong2022new,lin2024role}. Recently, the bias stress instability study of metal-oxide semiconductor attracts a lot of attentions with process optimization such as capping layers\cite{lin2024enhancement,lin2024role}, annealing conditions\cite{kim2023thermally}, dopant aspect and concentration\cite{wang2024ge,kim2023demonstration}, or plasma treatment\cite{yang2024igzo}. In terms of environmental sensitivity, poly-Si demonstrates a greater stability, as its covalent bonds and crystalline structure make it less reactive to humidity and oxygen, whereas metal-oxide semiconductors are highly sensitive to moisture and oxygen exposure, which can alter their electrical properties. This environmental sensitivity often necessitates encapsulation for metal-oxide semiconductors\cite{chowdhury2015effect}, while poly-Si remains more robust in uncontrolled environments despite its limitations in mobility and thermal budget requirements. The thermal stability of ferroelectric materials is another crucial consideration in sequential stacking, as the bottom tiers are subjected to additional thermal cycles during processing\cite{kim2022design}. 

\begin{figurehere}
   \centering
    \includegraphics[scale=0.1,width=\textwidth]{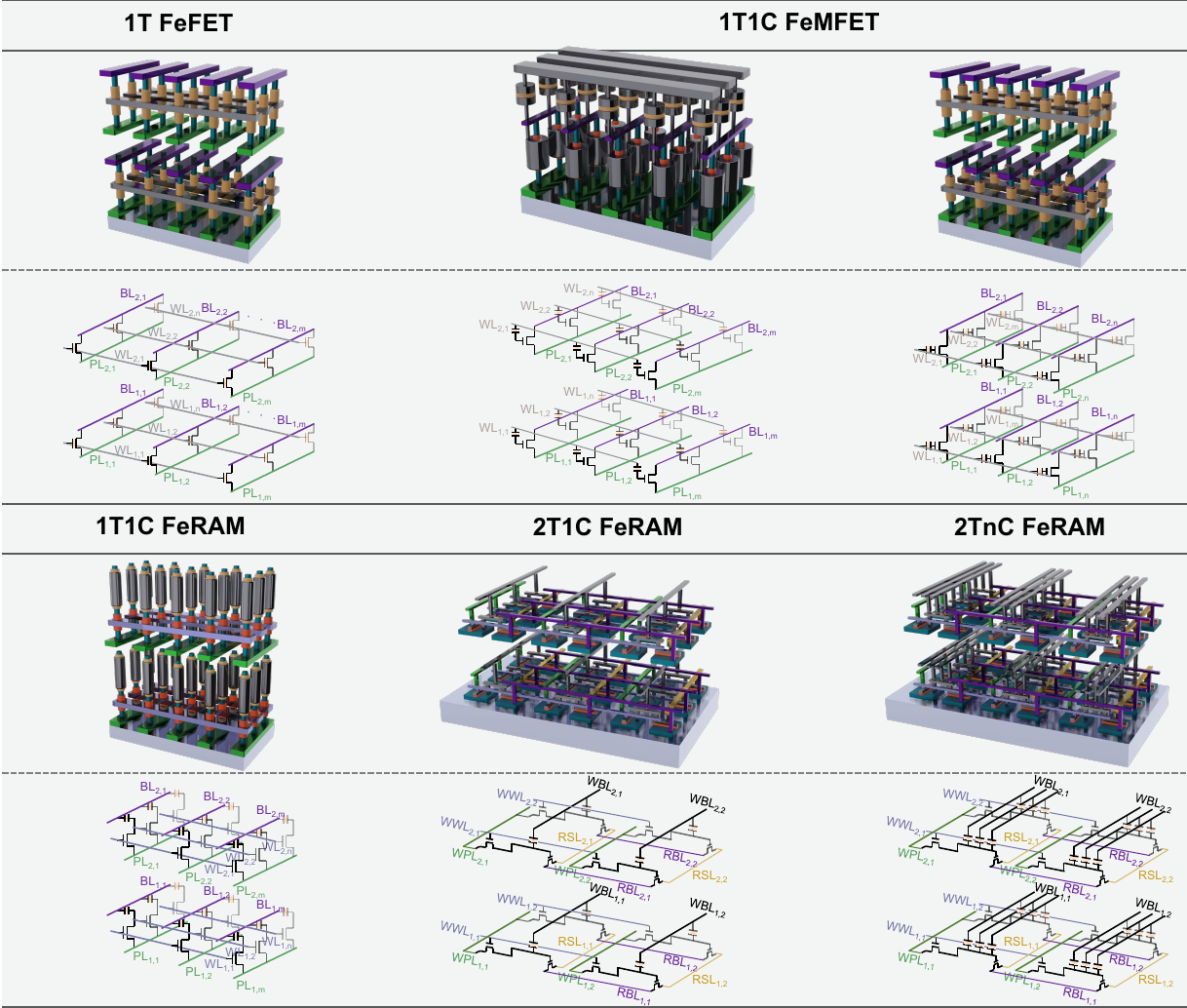}
    \captionsetup{parbox=none} 
    \caption{\textit{\textbf{3D sequential stacking of ferroelectric memories.} All these ferroelectric memories can be integrated sequentially in 3D through layer by layer stacking. But it is not bit-cost scalable because the processing cost increases almost linearly with the number of layers, which also introduces the thermal budget issue as the bottom tiers suffer more thermal process than top tiers.}}
    \label{fig:Sequential Stacking}
\end{figurehere}

The 3D sequential stacking for a 1T FeFET memory array can be implemented as shown in Fig.\ref{fig:Sequential Stacking}, where planar or vertical gate-all-around (VC-GAA) FeFETs are stacked layer by layer in a sequential manner. In this design, metal-oxide semiconductor channel is heavily investigated due to its low deposition temperature and flexible deposition methods including sputtering and also atomic-layer deposition\cite{sun2021first,chen2022first,lin2021high}. A major benefit of using metal-oxide semiconductor channel is to avoid the formation of a defective interfacial layer between the semiconductor and the HfO\textsubscript{2}-based ferroelectric layer, which plays a major role in contributing to endurance degradation in Si-based FeFETs\cite{gong2017study,zagni2023reliability}. However, the erase operation challenges remain for metal-oxide semiconductor channel FeFETs due to the lack of holes, and it could be improved by device scaling which enhances the source/drain fringe field to erase polarization and introducing the intermediate oxygen-deficient layer with a higher density of oxygen vacancies as mentioned in parallel 3D Fe-NAND. Targeted at embedded applications, rapid advancements of BEOL FeFETs have been demonstrated towards low operation voltage through stack thinning and material optimization, high endurance through interface optimization, and highly scaled dimensions\cite{sun2021first,kirtania2024amorphous,datta2024amorphous,chen2022first}. As an example, two-tier structure consists of two 10x10 FeFET sub-arrays using W doped In\textsubscript{2}O\textsubscript{3} (IWO) channel is reported to serve as the TCAM cell\cite{dutta2021lifelong} with record low voltage of 1.6 V and also the in-situ Hamming distance computation ability, which indicts that BEOL sequential stacking FeFET array is suitable for embedded compute in memory (CiM) applications.

As for 3D sequential stacking 1T-1C FeMFET, two design options are available, depending on the orientation of the ferroelectric capacitor, as introduced in parallel integration. If the MFM capacitor is formed after the VC-GAA MOSFET\cite{lee2022vertical}, the area of the MFM capacitor can be defined easily by lithography (first 1T-1C FeMFET design in Fig.\ref{fig:Sequential Stacking}), this will introduce another MFM capacitor to transistor pitch between two cells in horizontal direction and consequently reduce the array density, even though the area ratio between MFM and MOSFET can be tuned without changing the MFM capacitor area because the MOSCAP area can be adjusted by changing the channel height. The second 1T-1C FeMFET design in Fig.\ref{fig:Sequential Stacking} shows the architecture where the orientation of MFM capacitor is the same as the gate of the MOSFET. Here, the area of MFM capacitor is defined by the horizontal recess length determined by the horizontal etch time and rate, which is not that precise but could increase the density as no additional photolithography step is required. Additionally like sequential FeFET, two layer FeMFET has also been reported to be usable for TCAM application\cite{joh2023ferroelectric}.

Sequential stacking of 1T-1C FeRAM can be done by repeating the planar array\cite{10185243,shi2024record}, similar to the 4F\textsuperscript{2} DRAM, which consists of a VC-GAA MOSFET and a vertical cylindrical ferroelectric capacitor to increase the surface area (Fig.\ref{fig:Sequential Stacking}). A major milestone has been reported recently that achieves 32Gb capacity\cite{ramaswamy2023nvdram}, which presents dual layer sequential stacking 1T-1C FeRAM with dual gate poly-Si channel access transistor, which are fabricated above the CMOS circuitry. It also shows an almost unlimited endurance and also retention at different temperatures. In the case of 2T-1C FeRAM, the two transistors are typically arranged to provide enhanced control over the read and write operations, improving the overall reliability and performance of the memory cell, while sacrificing density. Integrating multiple MFM capacitor in 2T-nC FeRAM with one pair of write and read transistors could offset the density loss. In all these sequential stacking 3D designs, material choices and thermal processes should be considered carefully due to the layer by layer stacking.

\section*{\textcolor{sared}{\large Outlook}}

At the end, a summary of different types of ferroelectric memories and their 3D integration is shown in Fig.\ref{fig:Outlook}.
Parallel stacking NAND FeFET offers very high density and its process is highly compatible with NAND flash. Moreover, due to the W/L ratio dependent sense margin, the area of FeFET could keep shrinking down given manageable device variability. It is widely used for information storage applications. Additionally, current-domain CiM designs using the Fe-NAND structure have been explored, leveraging the extremely high density\cite{kim2023highly,zhao2023situ,shim2020technological,jin2023multi}. However, there are two issues limiting its application. One is that with the number of stacked layers of Fe-NAND increases, the sensing current is degraded due to series resistance from the vertical channel. The other is that there is not enough memory window for reliable multi-bit storage. Recently, tmany works have focused on gate-side injection, which can significantly enlarge the memory window. However, a comprehensive evaluation of all reliability aspects is still required. In the AND FeFET array, FeFETs in a string are connected in parallel, so not all unselected transistors in the same string need to be turned on. This enables fast and random access, making it an attractive array choice for accelerating inference in large neural network models. It also benefits from good scalability due to the same W/L-dependent sense margin. The main challenges are from its complex integration process and conjected routing for WL, BL, and SL. 

\begin{figurehere}
   \centering
    \includegraphics[scale=0.1,width=\textwidth]{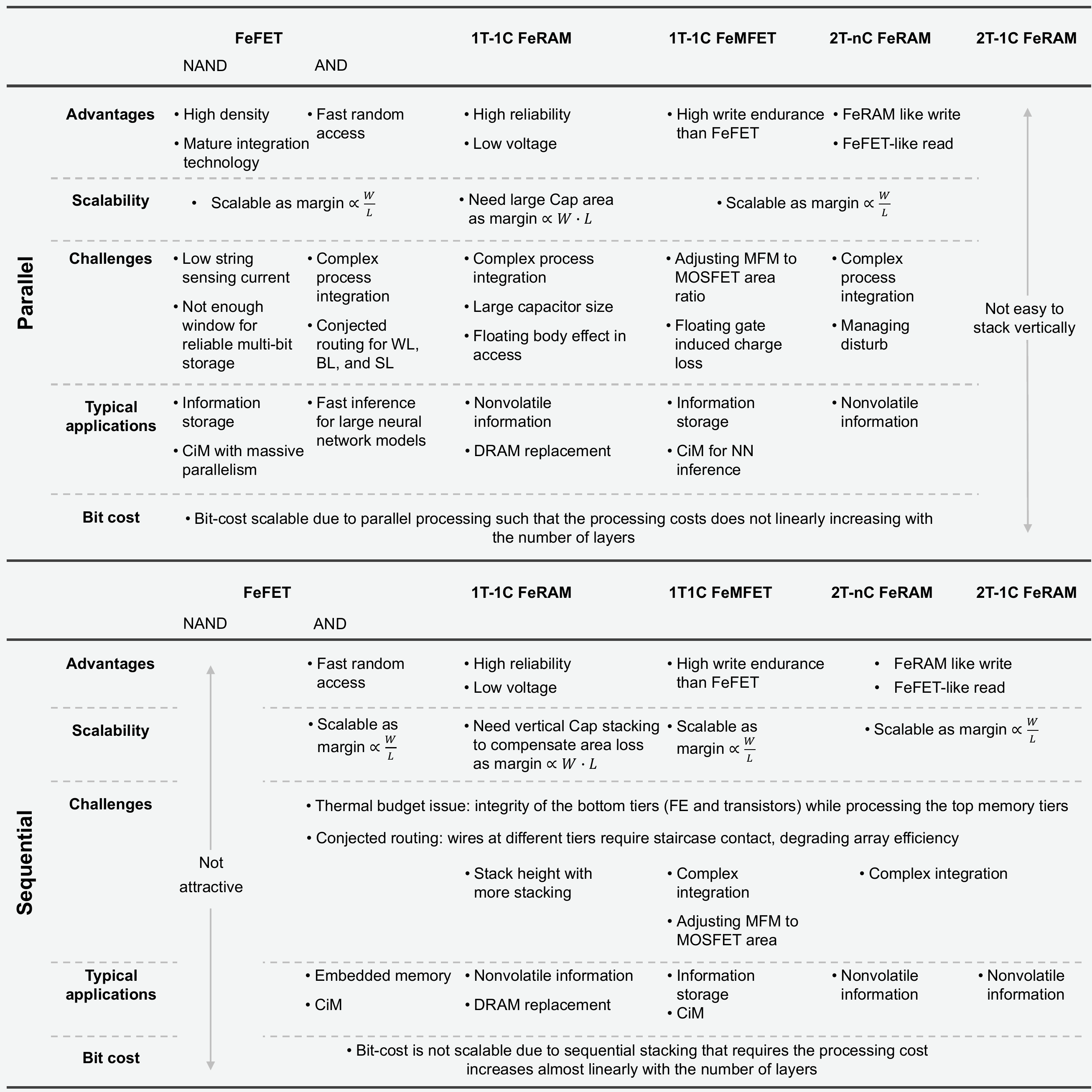}
    \captionsetup{parbox=none} 
    \caption{\textit{\textbf{Summary of parallel and sequential 3D integration of ferroelectric memories and their applications.} Parallel 3D stacking designs have the common advantages of high density and bit-cost scalability due to parallel processing such that the processing costs does not linearly increasing with the number of layers. The sequential 3D stacking designs have almost linearly increasing process cost with the number of stacked layers and also face the common challenges from the thermal budget issue and complex routing design.}}
    \label{fig:Outlook}
\end{figurehere}

For the parallel stacking 1T-1C FeRAM, it has the advantages of high reliability and low operation voltage. Though very challenging, this design, having the similar 1T-1C DRAM cell structure, is expected to replace the DRAM due to the similar structure, performance, and its non-volatility. Many of its challenges are also similar with 3D DRAM. Firstly, its scalability is bad, because the sense margin is proportional to the area of ferroelectric capacitor, so that a large capacitor is required. Secondly, the integration process is very complex. 
Lastly, whether the floating body effect in the access transistor affects the cell operation and reliability remains unclear. The parallel stacking 1T-1C FeMFET offers the possibility for tuning the area ratio between ferroelectric capacitor to MOSFET, which means adjusting the voltage distribution between ferroelectric capacitor and MOS capacitor. Thus, the MW could be enlarged and the charge trapping can be suppressed. This enables potentially high write endurance and lower write voltage compared with FeFET. So, it could also be used for non-volatile information storage and CiM for NN inference. However, the flexibility of adjusting the area ratio also introduces more complex process to define the ferroelectric capacitor. The other challenge is from the floating gate induced charge loss. For 2T-1C FeRAM, there are no easy approach of parallel 3D stacking. For 2T-nC FeRAM, it has the lower write voltage, large sense margin due to FeRAM-like write and FeFET-like read, respectively. It could also used for nonvolatile storage. The challenges limiting its application are complex integration process and disturb management. All these parallel stacking designs have bit-cost scalability due to the parallel process such that the process costs does not linearly increasing with number of stacked layers.

As mentioned before, sequential stacking designs have the common challenges including thermal budget issue and conjected routing. 
Sequential stacking AND FeFET has the common advantages from AND structure, that is fast random access, which makes it suitable for embedded memory and CiM applications. Additionally, it benefits from a large sense margin characteristic of FeFET memory cells. The sequential-stacked 1T-1C FeRAM maintains the same advantages as its parallel-stacked counterpart. However, since its sense margin depends on capacitor area, vertical capacitor stacking becomes necessary to compensate for area reduction in the horizontal plane. This creates a critical trade-off between maintaining adequate sense margin and managing process complexity in the vertical dimension.Sequential stacking 1T-1C FeMFET faces same challenges as the parallel 1T-1C FeMFET. Sequential stacking 2T-1C and 2T-nC FeRAM can also take a position in non-volatile storage with its fast and low power write and large sense margin. 

\section*{\textcolor{sared}{\large Conclusions}}
In conclusion, this review highlights the transformative potential of 3D ferroelectric memory, particularly utilizing HfO\textsubscript{2}-based materials, as a promising solution to overcome the limitations of traditional memory technologies. The comprehensive classification of polarization sensing mechanisms provides valuable insights into the design, performance, and scalability of ferroelectric memory architectures. By exploring various architectures within this framework, we have outlined their key advantages and trade-offs, offering a roadmap for optimizing sensing mechanisms to achieve improved read accuracy, power efficiency, and endurance. As we move towards next-generation memory solutions, this work underscores the importance of 3D integration and sensing enhancements in enabling more efficient and reliable nonvolatile memory technologies. Ultimately, this review lays the foundation for future advancements that can meet the demands of high-performance, low-power, and AI-driven computing applications, driving the evolution of memory technologies well into the future.

\bibliography{ref}
\bibliographystyle{naturemag}
\section*{\large Acknowledgments}

This work is primarily supported by the U.S. Department of Energy, Office of Science, Office of Basic Energy Sciences Energy Frontier Research Centers program under Award Number DESC0021118. It is also partially supported by SUPREME and PRISM centers, two of the SRC/DARPA JUMP 2.0 centers.

\section*{\large Author contributions}

K.N. and G.X. proposed the project. J.D. led the project. A.K. and V.N. contributed to the discussions. All authors contributed to write up of the manuscript.

\section*{\large Competing interests}
The authors declare that they have no competing interests.

\end{document}